\documentclass[manuscript]{aastex631}
\submitjournal{PSJ}

\graphicspath{{./}{figures/}}

\begin{document}
\title{Methane-saturated layers limit the observability of impact craters on Titan}

\correspondingauthor{Shigeru Wakita}
\email{shigeru@mit.edu}

\author[0000-0002-3161-3454]{Shigeru Wakita}
\affiliation{Department of Earth, Atmospheric, and Planetary Sciences, Purdue University, West Lafayette, IN, USA}
\affiliation{Department of Earth, Atmospheric and Planetary Sciences, Massachusetts Institute of Technology, Cambridge, MA, USA}

\author[0000-0002-4267-093X]{Brandon C. Johnson}
\affiliation{Department of Earth, Atmospheric, and Planetary Sciences, Purdue University, West Lafayette, IN, USA}
\affiliation{Department of Physics and Astronomy, Purdue University, West Lafayette, IN, USA}

\author[0000-0003-3715-6407]{Jason M. Soderblom}
\affiliation{Department of Earth, Atmospheric and Planetary Sciences, Massachusetts Institute of Technology, Cambridge, MA, USA}

\author[0000-0002-7535-663X]{Jahnavi Shah}
\affiliation{Department of Earth Sciences, The University of Western Ontario, London, ON, Canada}

\author[0000-0003-3254-8348]{Catherine D. Neish}
\affiliation{Department of Earth Sciences, The University of Western Ontario, London, ON, Canada}
\affiliation{The Planetary Science Institute, Tucson, AZ, USA}



\begin{abstract}
As the only icy satellite with a thick atmosphere and liquids on its surface, Titan represents a unique end-member to study the impact cratering process.
Unlike craters on other Saturnian satellites, Titan's craters are preferentially located in high-elevation regions near the equator. 
This led to the hypothesis that the presence of liquid methane in Titan's lowlands affects crater morphology, making them difficult to identify. 
This is because surfaces covered by weak fluid-saturated sediment limit the topographic expression of impact craters, as sediment moves into the crater cavity shortly after formation.
Here we simulate crater-forming impacts on Titan's surface, exploring how a methane-saturated layer overlying a methane-clathrate layer affects crater formation. 
Our numerical results show that impacts form smaller craters in a methane-clathrate basement than a water-ice basement, due to the differences in strength. 
We find that the addition of a methane-saturated layer atop this basement reduces crater depths and influences crater morphology.
The morphology of impact craters formed in a thin methane-saturated layer are similar to those in a "dry" target, but a thick saturated layer produces an impact structure with little to no topography. 
A thick methane-saturated layer (thicker than 40\% of the impactor diameter) could explain the dearth of craters in the low-elevation regions on Titan.
\end{abstract}



\section{Introduction} \label{sec:intro}
Saturn’s largest moon, Titan, is composed of an icy shell overlying a liquid water ocean \citep{Sotin:2021}. Titan is unique amongst the icy satellites, as it possesses a thick, nitrogen--methane atmosphere.  The presence of an atmosphere has led to the modification of its surface, including the erosion and weathering of its impact craters \citep{Neish:2013, Neish:2016, Hedgepeth:2020}.
Though impact craters are found on Titan, there is a dearth of craters compared to other icy moons. 
Recent analyses of Cassini spacecraft observations have revealed only 90 possible impact craters \citep{Hedgepeth:2020}, 
which range from 3 km to 400 km in diameter. 
Craters are categorized from certain to possible depending on their morphologic features \citep{Wood:2010,Hedgepeth:2020}; for example, 
a circularly shaped depression with a clear rim and floor is categorized as a certain crater whereas a circular shape without a clear rim and floor is categorized as a possible crater. 
Erosional processes could explain why it is more difficult to recognize these \textit{possible} craters. 
Some other processes, however, might also be affecting the observed morphology of Titan's craters. 
Even when accounting for observational coverage bias, Titan's craters are distributed inhomogeneously across its surface, with more craters in the highlands than lowlands \citep{Neish:2014}. 

In addition to global-scale trends in crater density, craters on Titan are typically shallower than craters on Jupiter’s comparably sized icy moons, Ganymede and Callisto, which are thought to be the best “airless” analogues to Titan \citep{Hedgepeth:2020, Neish:2013}.
The moons primarily differ because of the presence of a large methane--nitrogen atmosphere on Titan, which allows for erosion and weathering of the surface. 
Specifically, Titan’s atmosphere enables the movement of the sand, which can result in the aeolian infill of its craters \citep{Lorenz:2006, Neish:2013}. 
Fluvial erosion likely also plays a role in the shallowing of Titan’s craters, as methane rainfall allows for overland flow that leads to bedrock incision \citep{Neish:2016, Hedgepeth:2020}. This can contribute to the erosion of the crater, degrading the rim \citep{Hedgepeth:2020, Neish:2013} and erasing the crater morphology in a few hundred Myrs \citep{Neish:2016}. 
These processes may provide at least part of the explanation for the anomalous depths of Titan’s craters. 
As such, liquid methane plays an important role in the evolution of Titan's impact craters, and fluvial erosion in the polar region may help to explain the dearth of craters there.
However, it is unlikely to explain the dearth of craters in the equatorial and mid-latitude lowlands, where rainfall is predicted to be less than that at the poles \citep{Turtle:2018a}. 
While methane rain could erode these craters, there is no reason to expect significantly different amounts of rain at different elevations. 

In these lowland regions, the unique presence of liquids on Titan's surface may help to explain the absence of impact craters. 
When an impactor hits the surface of the terrestrial body, it can form a crater \citep{Melosh:1989}.
An impact produces a shockwave and subsequent rarefaction or release wave which propagates and engulfs the target material and sets up an excavation flow. 
This flow forms a bowl shaped transient crater that subsequently collapses under the force of gravity producing the final crater morphology. 
The excavation stage of cratering has some sensitivity to target strength, but the collapse of the crater in the modification stage is very sensitive to target strength \citep[e.g.,][]{Ivanov:2010, Miljkovic:2016, Johnson:2018a}. 
Here we explore the effect that variations in target strength might have on the morphology of Titan’s craters, including the possibility that craters formed in weak material saturated in liquid methane are undetectable from Cassini observations, or are very easy to subsequently erase through erosion and weathering. 

In this work, we focus on the hypothesis that Titan’s lowlands are weakened by the presence of liquids preferentially located at low elevations, resulting in a deficit of craters these elevations \citep{Neish:2014}. 
At present day, methane lakes and seas are observed within Titan's polar regions \citep{Hayes:2008, Stofan:2007}, but in the past, Titan may have had a near-global methane/ethane ocean \citep{Burr:2013, Lunine:1983, Larsson:2013}. 
This liquid may soak into Titan’s water-ice crust forming a “methane-saturated” surface layer that would weaken its icy crust. 
This would, in turn, enhance collapse during the modification stage of the impact cratering process \citep[e.g.,][]{Melosh:1989}, similar to the formation of the Chesapeake Bay and other “marine” impact craters on Earth \citep{Collins:2005}. 
Thus, if there is a “methane-saturated” layer that is concentrated in the lowest elevations on Titan, its presence could explain the limited number of lowland craters \citep{Neish:2014}.  

An additional consideration, however, is the possible presence of clathrate hydrates. Liquid methane (and ethane) may also diffuse into the ice that comprises Titan's crust and form methane (or ethane) clathrate hydrates \citep{Choukroun:2010}. 
Both methane clathrate hydrate and ethane clathrate hydrate are stable at the temperatures and pressures expected in Titan surface crustal layers \citep{Choukroun:2010, Vu:2020} and, based on recent experiments conducted using liquid ethane, should form on very short timescales (about 1 Earth year at 90 K \citep{Vu:2020}). 
If a methane-saturated layer is present, a methane-clathrate crustal layer is therefore also likely. 
Models of the thermal evolution \citep{Kalousova:2020} and density structure \citep{Cadek:2021} of Titan’s crust both support the presence of a methane-clathrate crust layer (possibly 5--10 km thick). 
Despite predictions of these layers, the roles of a methane-saturated layer and/or methane-clathrate layer on the cratering process have not been studied before.  

In this paper, we examine the influence that methane has on Titan’s craters. 
We first describe numerical methods for impact simulations, including the development of a strength model for methane clathrate. 
We next show our results for the resultant crater morphologies, obtained by varying the thickness of the methane-saturated and/or methane-clathrate layer and the impactor size. 
Finally, we discuss the effects of these methane layers on Titan’s impact craters. 

\section{Methods} \label{sec:meth}
In this paper, we numerically investigate the influence that methane-saturated layers and methane-clathrate layers have on crater formation. 
To accomplish this, we perform impact simulations using iSALE-2D, a shock physics code \citep{Collins:2016, Wunnemann:2006} based on the SALE hydrocode \citep{Amsden:1980} that has been developed to model planetary impacts and cratering. 
The code has been improved from the SALE code by including various equations of state and a strength model \citep{Collins:2004, Ivanov:1997, Melosh:1992}. 
In our simulations, we assume the impactor is a spherical body composed of water ice to represent a comet-like composition, as Titan’s impactors are thought to come from the outer solar system \citep{Zahnle:2003}. 
Note that Titan’s thick atmosphere limits the impactor size that can hit the surface \citep{Artemieva:2003}; 
impactors larger than $\sim$ 5 km in diameter can reach the surface virtually unaffected by the atmosphere, and the atmosphere has only a modest effect on impactors 1--5 km in diameter. 
Therefore, we consider impactor diameters ($D_{\rm imp}$) of 1, 2, 3, and 5 km. We fix the impact velocity at 10.5 km/s, which is the average impact velocity on Titan \citep{Zahnle:2003}. 
To resolve a layer in the target of a simulation requires at least 10 cells across that layer, e.g., 100 m for a 1-km thick layer, or 50 m for 0.5-km thick layer.
As we vary the thickness of the methane-saturated layer according to the impactor size (see below), 
we choose a spatial resolution for the calculation region of 100 m for $D_{\rm imp}$ = 5 km and 50 m for the other sizes. 

We consider three possible target materials in our impact simulations: water ice, methane clathrate, and methane-saturated ice. 
For these three materials, we use the Tillotson equation of state (EOS) for water ice \citep{Tillotson:1962, Ivanov:2002}. 
This is because there is no shock physics EOS or Hugoniot data for methane clathrate or methane-saturated ice. 
We expect that the volume change associated with the crushing of clathrate cages would be qualitatively similar to the crushing of pore space. 
If correct, this would imply a more accurate EOS for clathrate would likely result in somewhat smaller transient crater diameters and somewhat larger transient crater depths \citep{Wunnemann:2006}. 
For the same size final crater, we would also expect a larger degree of impact heating and melting \citep{Wunnemann:2008}. 
Table \ref{tab:input} summarizes the material parameters used in our simulations. In order to focus on the effects of methane, we treat all materials as non-porous in this study. 
Below we describe how we model impacts into methane-clathrate and methane-saturated-ice layers.

\subsection{Methane clathrate}
There is currently no strength model of methane clathrate available for impact simulations, so we developed one based on experimental results. 
\citet{Durham:2003} examined the strength of methane clathrate at temperatures of 250--287 K and pressures of 50--100 MPa. 
Their experiments revealed that methane clathrate hydrate is 20--30 times stronger than water ice. 
Thus, we adopt the limiting strength of methane clathrate as 20 times greater than that of water ice (see Table \ref{tab:input}). 
Since the experimental temperatures are much higher than the conditions at Titan’s surface (94K), we extrapolated the experimental results to this lower temperature. 

\begin{figure}
\plotone{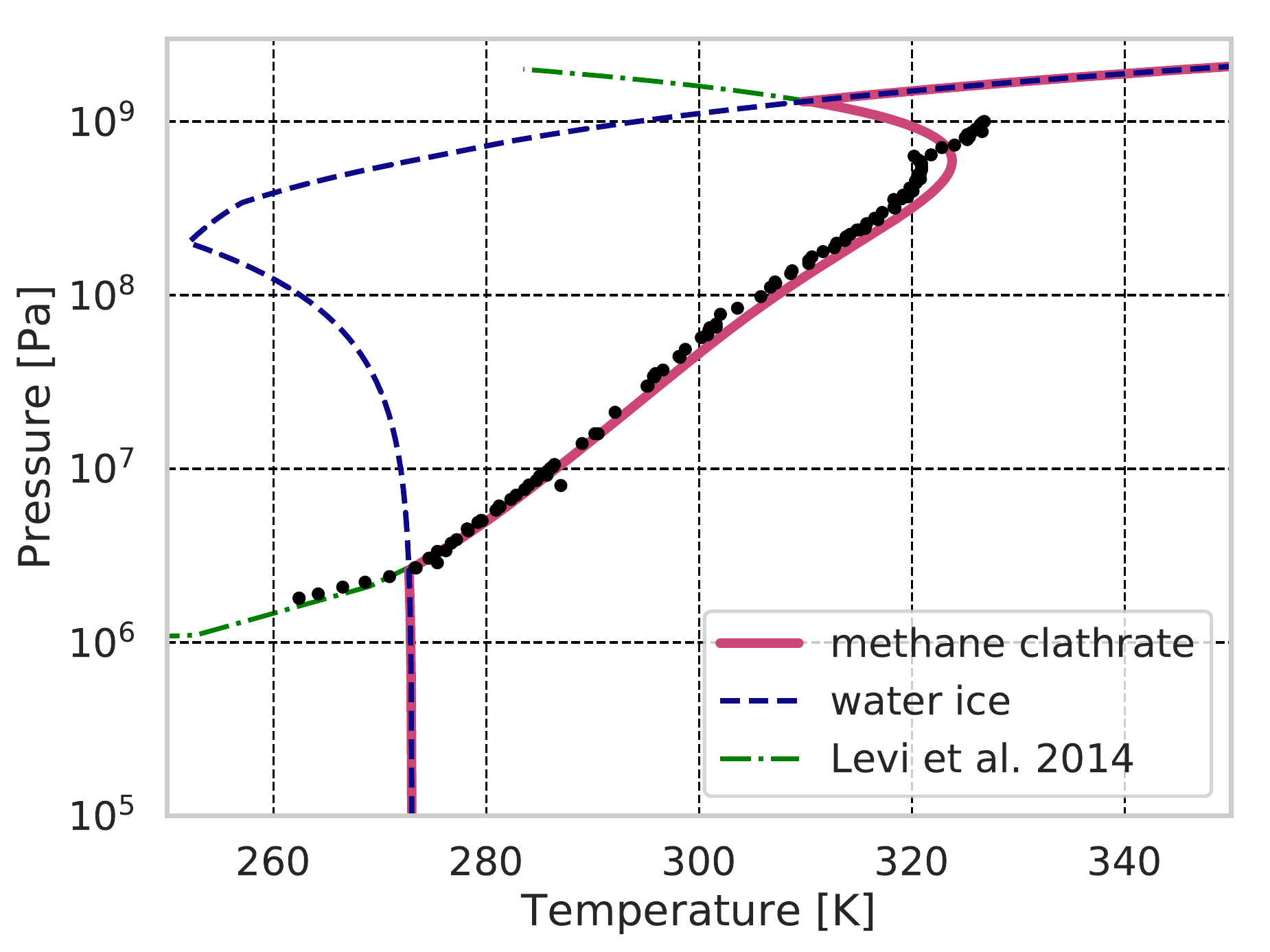}
\caption{
Melting temperature curve as a function of pressure. 
The pink solid line represents the melting temperature curve of methane clathrate in this study (see \S \ref{sec:meth}).
The melting temperature of water ice is shown as blue dashed line as a reference. 
The points depict experimental data \citep[][and references therein]{Sloan:2007}.
The green dash-dotted line are given by the equation in \citet{Levi:2014}.
\label{fig:tmelt}}
\end{figure}

\begin{figure}
\plotone{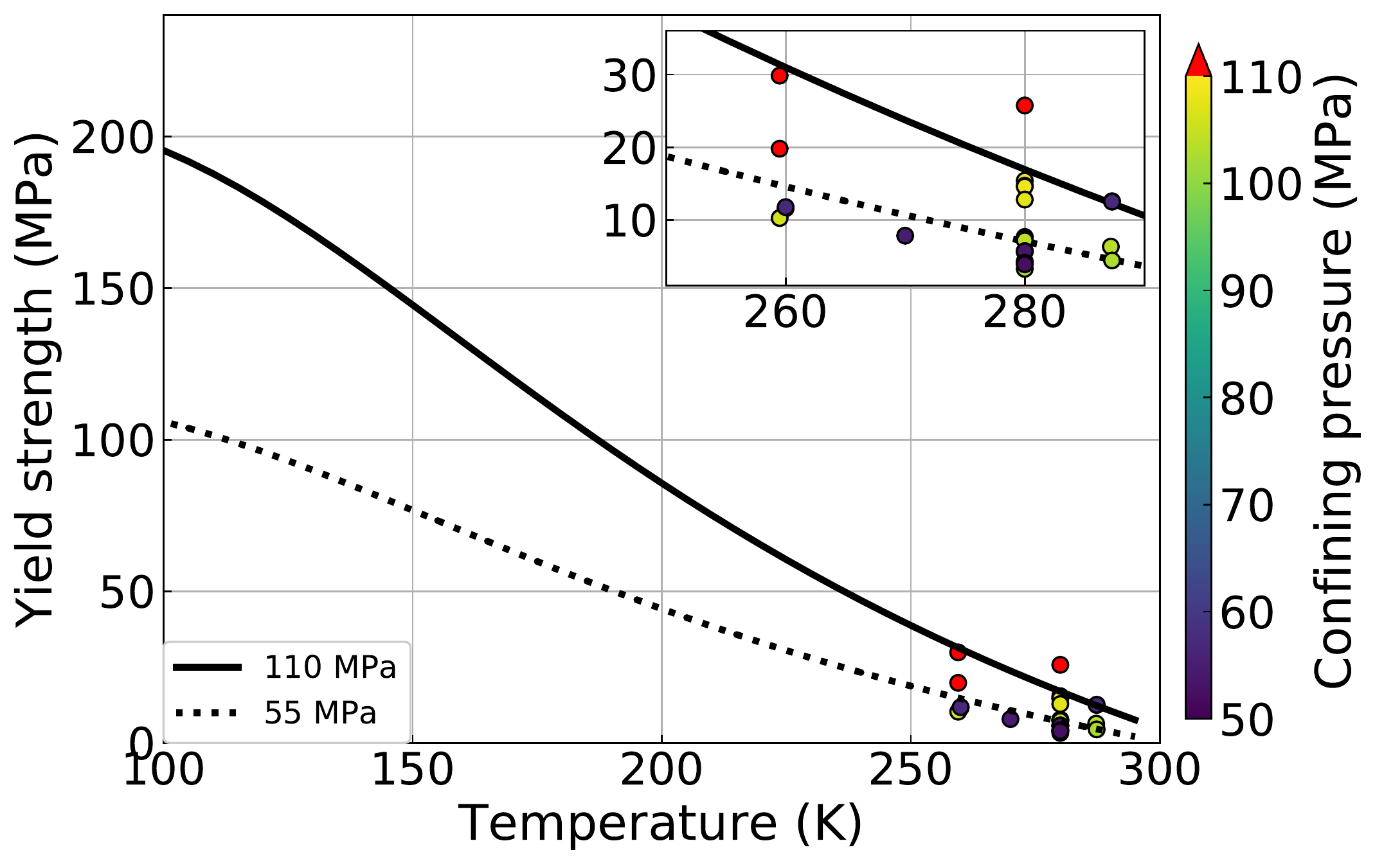}
\caption{
Yield strength of methane clathrate as a function of temperature. 
The symbols depict experimental data \citep{Durham:2003} colored by the confining pressure (MPa). 
Lines represent our strength model with the confining pressure of 100 MPa (solid line) and 50 MPa (dotted line).
\label{fig:strength}}
\end{figure}

The strength of the material depends on the temperature and approaches zero at the melting temperature. 
To describe this behavior, we use the melting temperature of methane clathrate and found a suitable value for the thermal softening parameter \citep{Ohnaka:1995}. 
As the melting temperature of methane clathrate depends on pressure \citep{Sloan:2007, Heidaryan:2019}, we model the melting temperature of methane clathrate as follows (see also Figure \ref{fig:tmelt}). 
At low pressures ($\lesssim3$ MPa), the melting temperature of methane clathrate is lower than that of water ice. 
As water ice is more stable than methane clathrate at these pressures, we assume that the melting temperature of methane clathrate is the same as that of water ice. 
At pressures of 3 MPa $\sim 1$ GPa, the melting temperature of methane clathrate is higher than that of water ice. 
Thus, we follow the melting temperature equation of methane clathrate in \citet{Levi:2014}, which well represents the experimental data \citep[e.g.,][and references therein]{Sloan:2007}. 
There are no experimental data for dissociation of methane clathrate at high pressure conditions ($\gtrsim 1$ GPa); 
methane clathrate might be less stable than water ice. 
Thus, we again use the melting temperature of water ice for methane clathrate for pressures $\gtrsim 1$ GPa.
Using our methane-clathrate melting temperature curve, we find that a thermal softening parameter of 0.8 represents the experimental data well (see below and Figure \ref{fig:strength}). 

Some parameters necessary to develop a strength model for methane clathrate (e.g., cohesion and frictional coefficient \citep{Bray:2014, Silber:2017}) cannot be derived from published experimental results. 
For these parameters, we use water-ice values. 
Note that at low temperatures, the cohesion and frictional coefficients of ice are similar to those of a variety of rocks \citep{Beeman:1988}. 
Thus, we expect this necessary assumption about cohesion and friction coefficients is a reasonable approximation. 
A comparison of our strength model to laboratory experiments is shown in Figure \ref{fig:strength}. 
Future experiments on the strength of methane clathrate at lower temperatures and higher stresses, however, are clearly desirable. 
\subsection{Methane-saturated ice}
While methane clathrate is stronger than water ice, methane-saturated ice is weaker than water ice. 
We model the methane-saturated layer as weak water-ice sediment. 
We use the same rheology as the water-saturated sediments in the Chesapeake Bay impact crater modelling \citep{Collins:2005}, and assume the limiting stress and frictional coefficient of the methane-saturated layer are lower than those of water ice (see Table \ref{tab:input}). 
The actual composition of the methane-saturated layer is unknown; 
the liquid methane may fill the solid (porous) water ice or unconsolidated water-ice sediment. 
Saturation by other liquid hydrocarbons (e.g., ethane) should have a similar weakening effect. 

\subsection{Layer thickness for three materials}
To evaluate the influence of the three materials (water ice, methane clathrate, and methane-saturated sediments), we change their thicknesses and relative placement to model plausible Titan crustal structures. 
Because methane clathrate is more insulating than water ice, thermal models predict a maximum clathrate thickness of $\sim$10 km \citep{Kalousova:2020}.
We first examine the case of methane clathrate over water ice, with a clathrate-layer thickness of 0--10 km and a $D_{\rm imp}$ = 5 km. 
Next, we consider models with 0 to 2 km of a methane-saturated layer on the top of a methane-clathrate basement ($l_{\rm CH_4(saturated)}$), with $D_{\rm imp}$ = 1, 2, 3, and 5 km. 
Models suggest the layer of porous-ice on Titan could be several hundred meters to a few kilometers thick \citep{Kossacki:1996}.
Since liquid hydrocarbon can diffuse into the pore space, this porous-ice layer may store hydrocarbons such as methane. 
In all cases, we assume a surface temperature of 94 K and a 3 K/km surface temperature gradient. 
This temperature gradient corresponds to a total ice-shell thickness of 40--100 km \citep{Beghin:2012, Mitri:2014, Tobie:2006}. 
In testing, we found that the presence of a liquid ocean beneath the icy crust has little effect on any craters formed by $D_{\rm imp} \leq $ 5 km, as long as the icy layer is at least 40 km thick. 
We, therefore, do not consider ice crust thickness as a variable and leave it fixed; we use a total ice crust thickness of $\sim$ 60 km in this work, though the exact value is not important so long as it is greater than $\sim$ 40 km.

\section{Results} \label{sec:res}
\subsection{Effect of a methane-clathrate layer} \label{subsec:clat}
\begin{figure}
\plotone{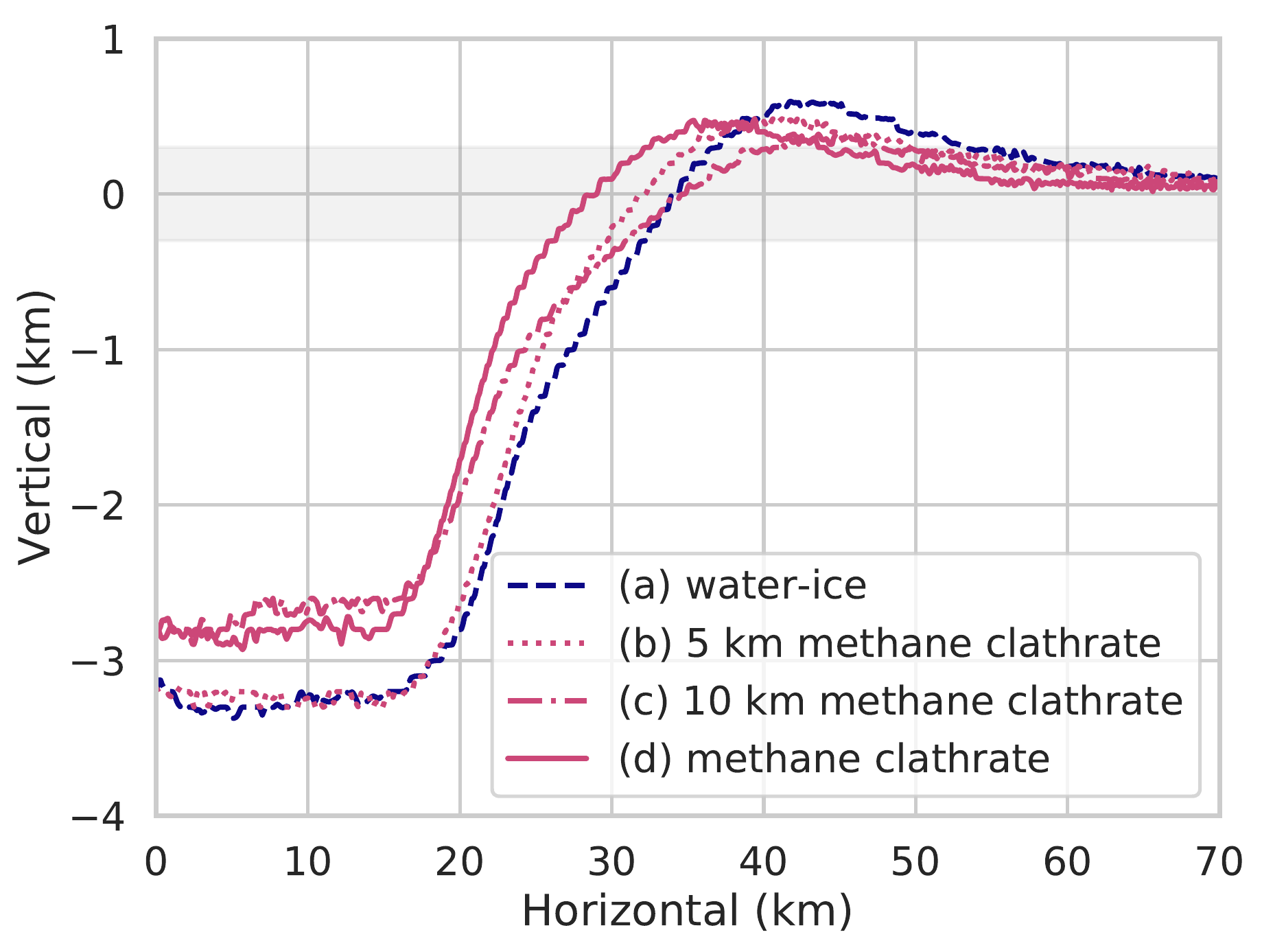}
\caption{
Surface profiles 2400 s after the impact of a 5-km-diameter impactor into different targets: (a) water-ice basement, (b) 5 km methane-clathrate layer over a water-ice basement, (c) 10 km methane-clathrate layer over a water-ice basement, and (d) methane-clathrate basement. The shaded region depicts the 300 m (3 cells) from the original target surface (6 cells in total height).
\label{fig:d5h2o}}
\end{figure}

We first compare crater formation between simulations with water-ice and methane-clathrate basements. 
As methane clathrate is stronger than water ice, the methane clathrate basement opens a smaller transient crater cavity than in the case of a water-ice target (see $t$ = 60 s in Fig. \ref{fig:d5_time}). This also limits the subsequent crater collapse (see supplemental movies \ref{sup:ch4} and \ref {sup:h2o}).
Figure \ref{fig:d5h2o} shows the surface profiles 2400 s after the impact of a 5 km-diameter impactor into: 
(a) a pure water-ice target, (b, c) a methane-clathrate layer over a water-ice basement, and (d) a pure methane-clathrate target. 
Using these results, we define the crater radius as the distance from the axis to the highest location (e.g., rim) and the depth as the height from the lowest location on the floor to the highest point on the rim. 
We find that craters formed in a pure methane-clathrate target are slightly smaller than those formed in a pure water-ice target (diameters and depths of $79\pm 0.6$ km and $3.4\pm 0.6$ km for methane clathrate vs $81\pm 0.6$ km and $4.0\pm 0.6$ km for water ice, for our 5-km-diameter-impactor case study). 
Because it is unclear if our resolution (100 m) is sufficient to fully resolve the crater rim, there is uncertainty in defining the crater radius and depth.
Impact simulation studies typically assume uncertainties of 2 or 3 cells; we conservatively take 3 cells for the uncertainty in this work, which results in errors in diameter and depth of 6 cells.

\begin{figure}
\plotone{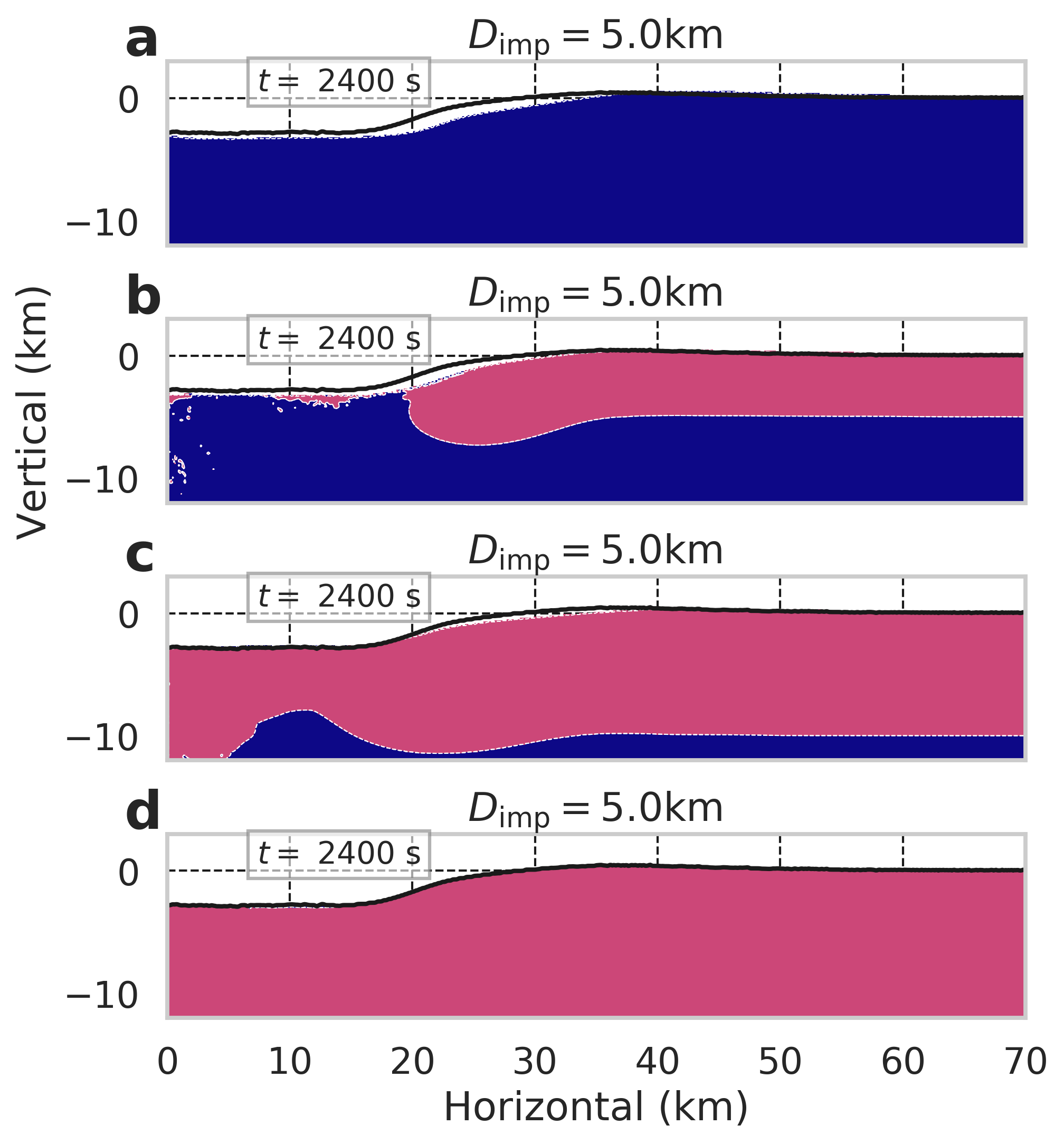}
\caption{
2D radial profiles of target composition, with water ice in blue and methane clathrate in pink, 2400 s after the impact of a 5 km-diameter impactor into different targets: (a) water-ice basement, (b) 5 km methane-clathrate layer over the water-ice basement, (c) 10 km methane-clathrate layer over a water-ice basement, and (d) methane-clathrate basement.
Black solid lines depict the surface profile of case (d) (the same the solid line in Fig. \ref{fig:d5h2o}).
\label{fig:d5h2o_line}}
\end{figure}

For the case of a methane-clathrate layer over a water-ice basement, we find that thicker methane-clathrate layers result in craters which topography approaches those formed in a pure methane-clathrate target. 
While a 5-km-thick methane-clathrate layer produces a crater with almost the same depth as the water-ice case (Fig. \ref{fig:d5h2o} (b)), a 10-km-thick methane-clathrate layer produces a crater that is more similar to the pure methane-clathrate case (Fig. \ref{fig:d5h2o} (c)). 
However, the composition within and surrounding the crater (e.g., crater floor, wall, and rim) differs depending on the thickness of the methane-clathrate layer (Fig. \ref{fig:d5h2o_line}). 
NASA's upcoming Dragonfly mission will reveal Titan’s surface in more detail \citep{Barnes:2021,Lorenz:2021}.
If future observations, such as those made by Dragonfly, detect any compositional differences between water ice and methane clathrate around an impact crater, we may be able to estimate the thickness of methane clathrate at the time of impact. 

\subsection{Effect of a methane-saturated layer} \label{subsec:sat}
\begin{figure}
\plotone{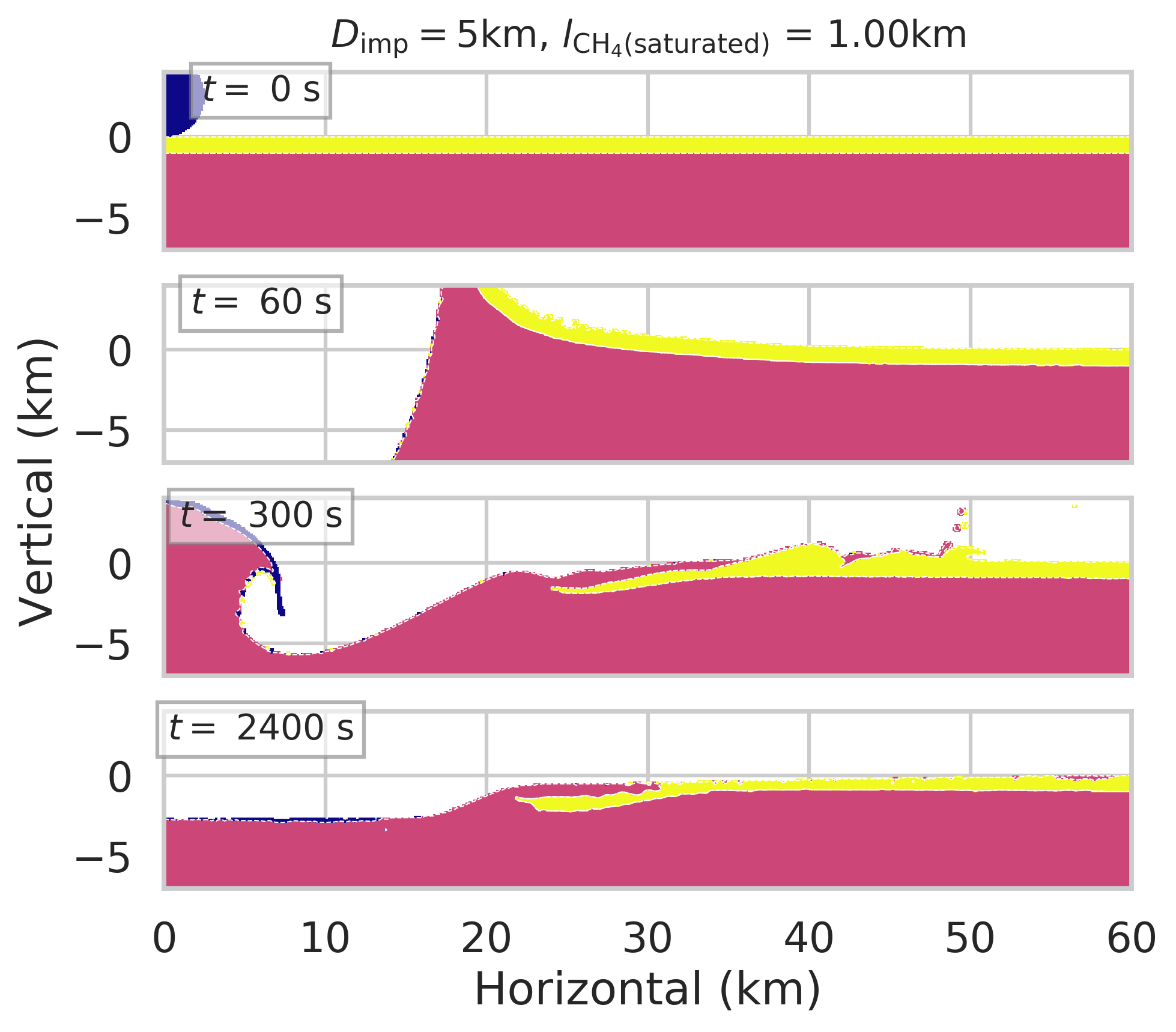}
\caption{
Time series for the impact of a $D_{\rm imp}$ = 5 km impactor into a target composed of a 1 km methane-saturated layer over a methane-clathrate basement. The blue color illustrates water ice (impactor), the pink color the methane clathrate, and the yellow color the methane-saturated layer. 
\label{fig:thin}}
\end{figure}

\begin{figure}
\plotone{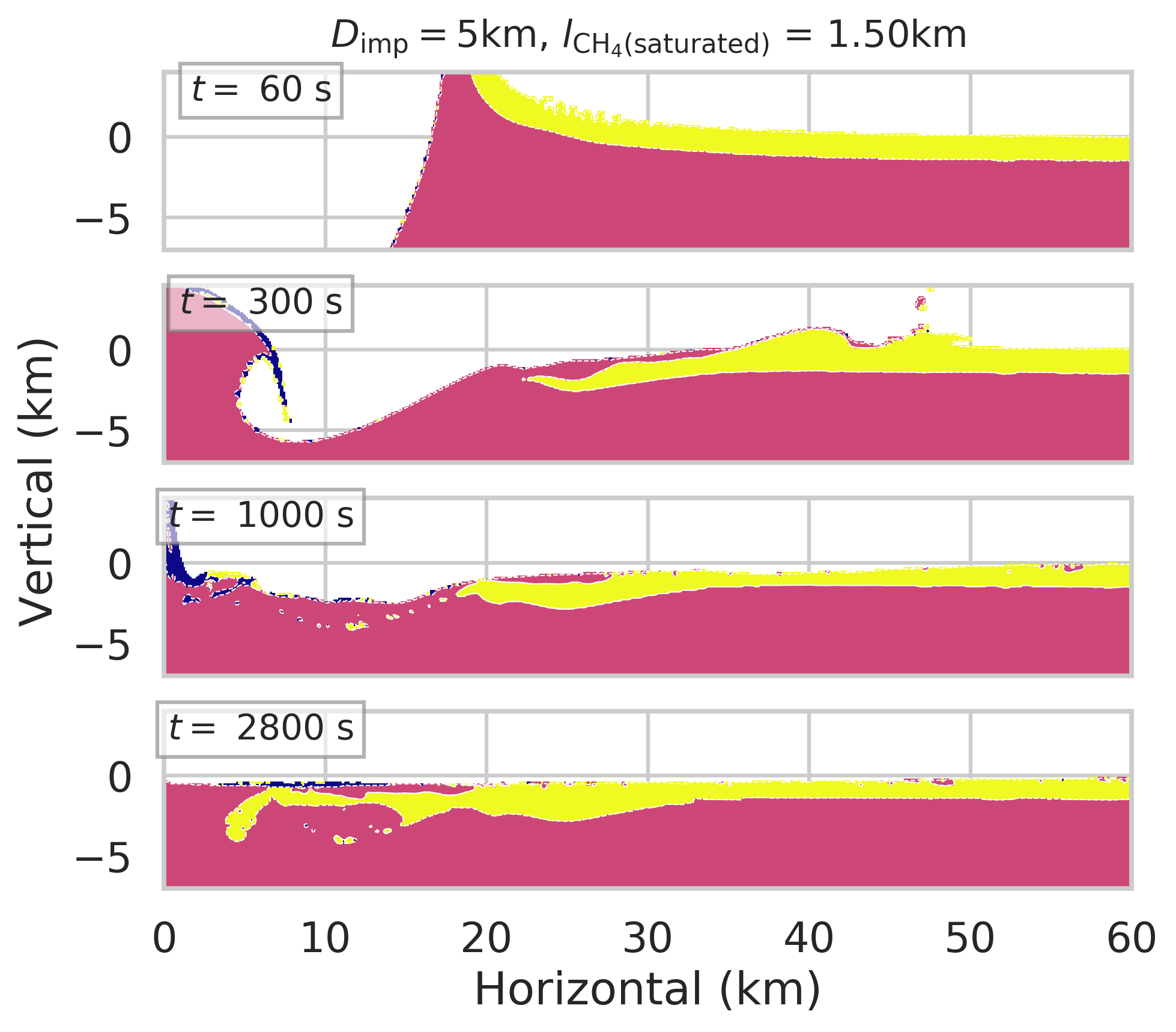}
\caption{
Same viewing geometry as Fig. \ref{fig:thin}, but for the case of a 1.5 km methane-saturated layer over a methane-clathrate basement with $D_{\rm imp}$ = 5 km.
\label{fig:interm}}
\end{figure}

\begin{figure}
\plotone{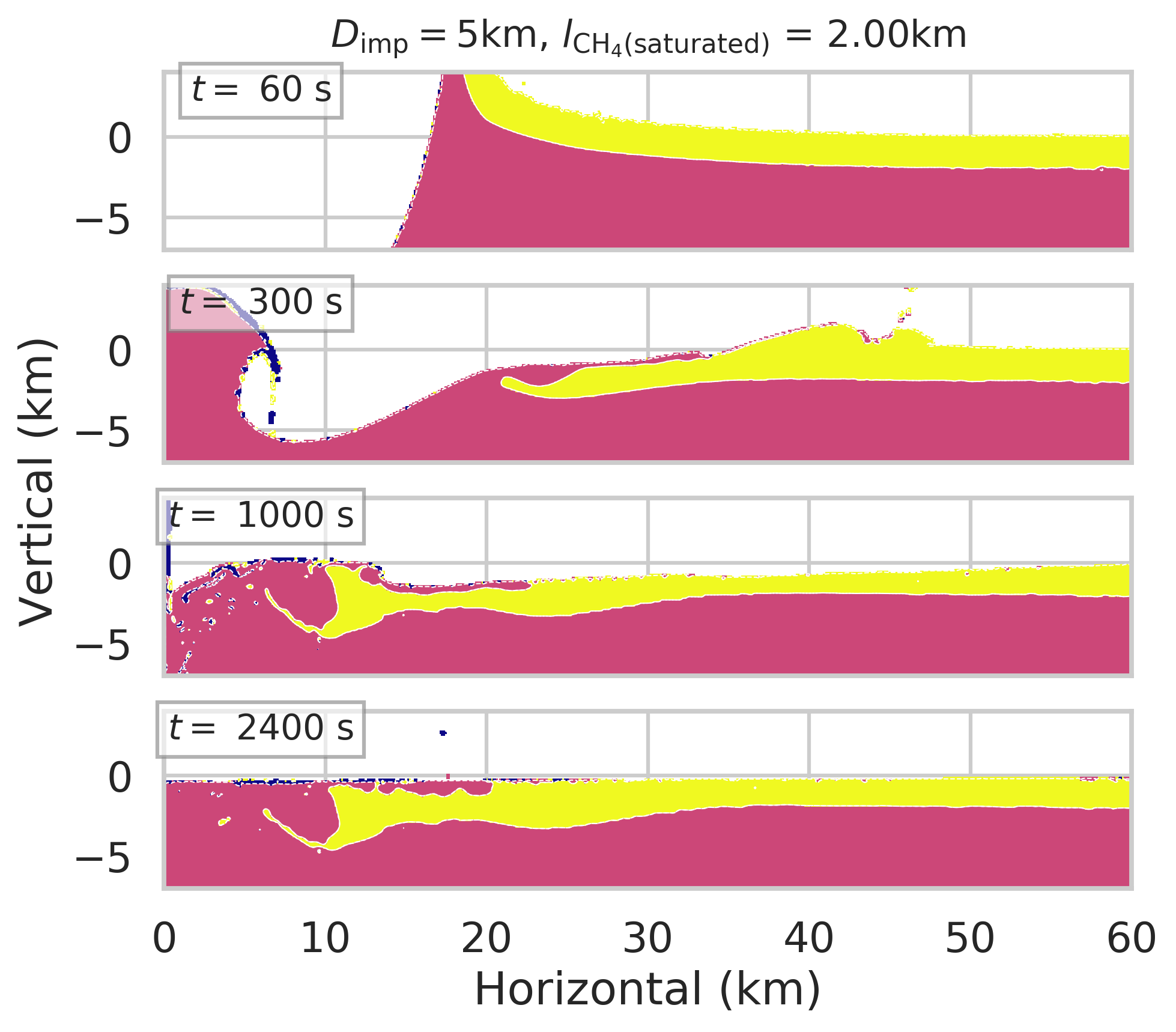}
\caption{
Same viewing geometry as Fig. \ref{fig:thin}, but for the case of a 2 km methane-saturated layer over a methane-clathrate basement with $D_{\rm imp}$ = 5 km.
\label{fig:thick}}
\end{figure}

\begin{figure}
\plotone{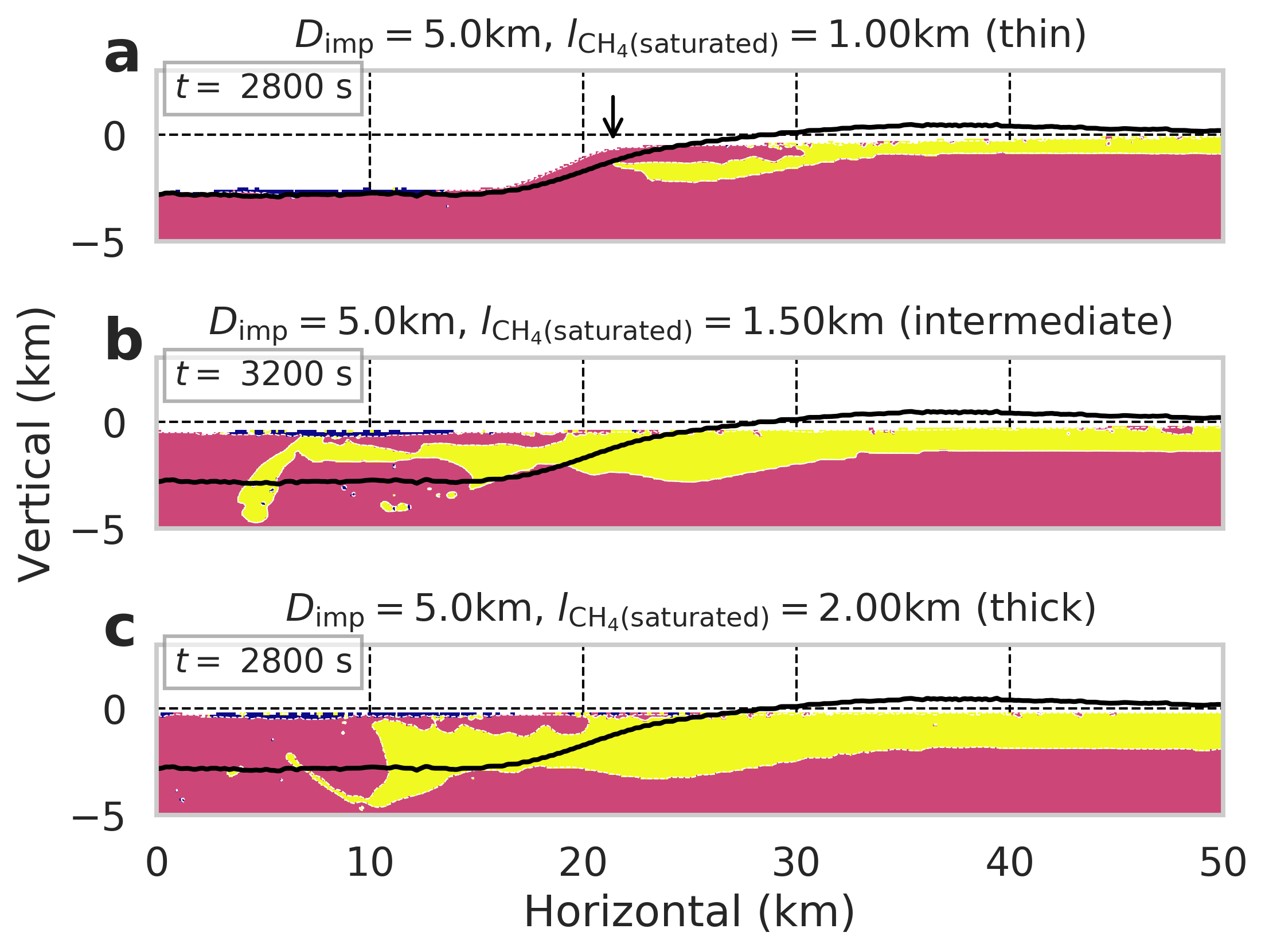}
\caption{
2D radial profiles of target composition for $D_{\rm imp}$ = 5 km impactors into a methane-saturated layer. 
Methane clathrate is shown in pink, methane-saturated ice in yellow, and water ice (impactor) in blue. 
Black solid lines depict the surface profile without the methane-saturated layer (i.e., into a pure methane-clathrate basement).
The thickness of the methane-saturated layer is (a) 1 km, (b) 1.5 km, and (c) 2 km. 
An arrow in panel (a) indicates the edge of the wall (see text). 
\label{fig:dimp5km}}
\end{figure}

The morphology of craters formed in a target covered by a methane-saturated layer depends strongly on the thickness of the saturated layer. 
First, we show time-series snapshots of 5 km-diameter-impactors into 1 km, 1.5 km, and 2 km thick methane-saturated layers (Figures \ref{fig:thin}, \ref{fig:interm}, and \ref{fig:thick}, as well as supplemental movies \ref{sup:ch4sthin}, \ref{sup:ch4smid}, and \ref{sup:ch4sthick}). 
Approximately 60 s after the initial impact, a crater is opened, which the methane-saturated ice moves towards and into. When the methane-saturated layer is 1.0 km thick (Fig. \ref{fig:thin}), the inward movement almost ceases at $t$ = 300 s, and the impact crater remains clearly distinguishable after formation ($t$ = 2400 s in Fig. \ref{fig:thin}). 
When the methane-saturated layer is 1.5 km thick, the methane-saturated material behaves similarly for the first 300 s (see Figs. \ref{fig:thin} and \ref{fig:interm}). 
The inward movement of this thicker saturated layer continues, however, ($t$ = 1000 s in Fig. \ref{fig:interm}) until this material eventually covers the impact cavity, leaving only a slight depression ($t$ = 2800 s in Fig. \ref{fig:interm}). 
For the case of a 2 km thick saturated layer, the methane-saturated material moves more dynamically and quickly fills the crater ($t$ = 300 s and $t$ = 1000 s in Fig. \ref{fig:thick}). 
As a result of this infill, the final crater has no topographic expression ($t$ = 2400 s in Fig. \ref{fig:thick}). 
Figure \ref{fig:dimp5km} summarizes the crater profile of the $D_{\rm imp}$ = 5 km cases with the methane-saturated layer ranging from 1 to 2 km in thickness. 
When there is a methane-saturated layer, the topographic expression of crater differs from that without a methane-saturated layer (see black lines in Fig. \ref{fig:dimp5km}). 
Therefore, the saturated-layer thickness has a strong control on the final crater morphology and whether a crater is easily recognizable from an orbital platform like Cassini. 

\begin{figure}
\plotone{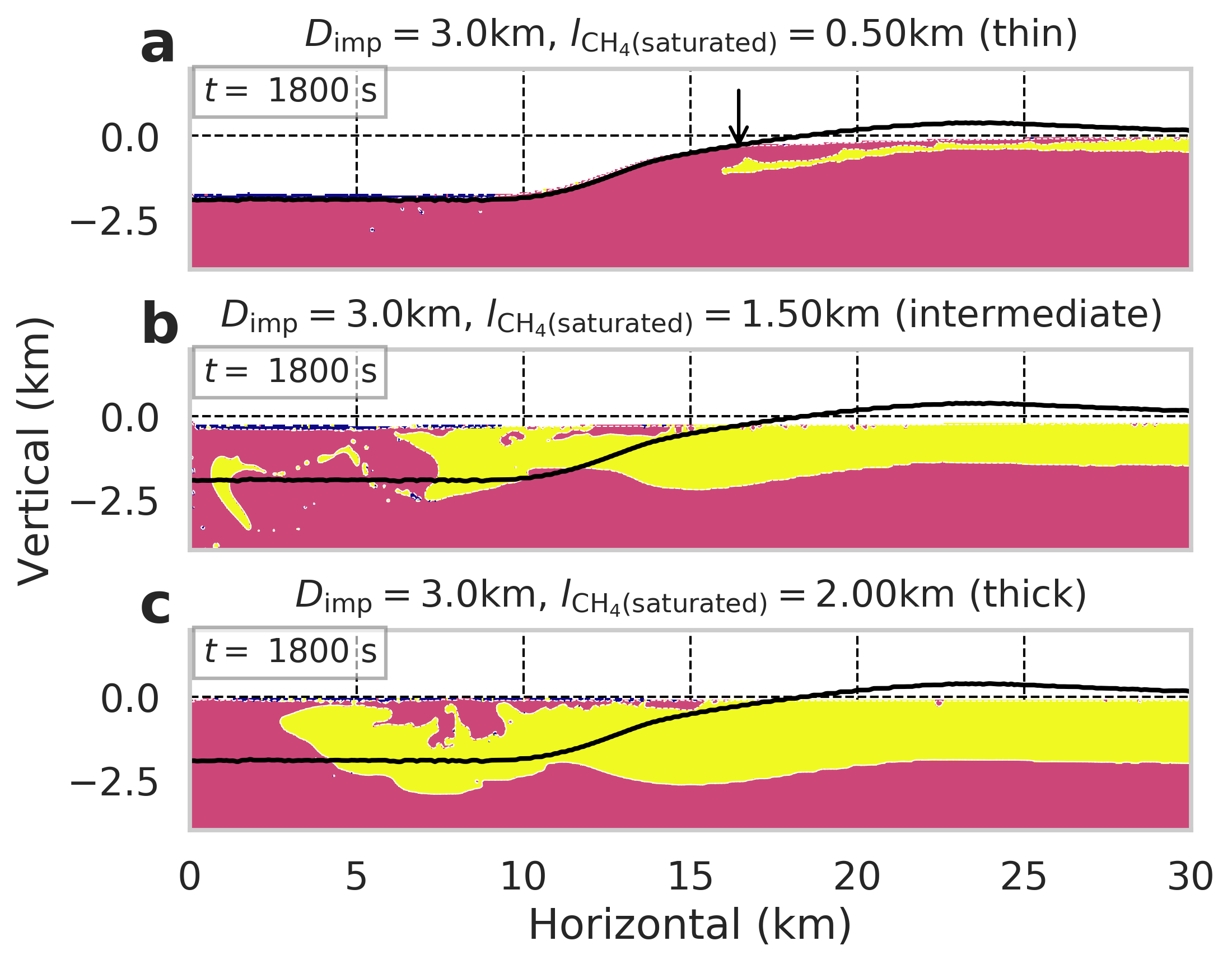}
\caption{
Same viewing as Fig. \ref{fig:dimp5km}, but for the case of $D_{\rm imp}$ = 3 km. The thickness of the methane-saturated layer is (a) 0.5 km, (b) 1.5 km, and (c) 2 km. 
\label{fig:dimp3km}}
\end{figure}

\begin{figure}
\plotone{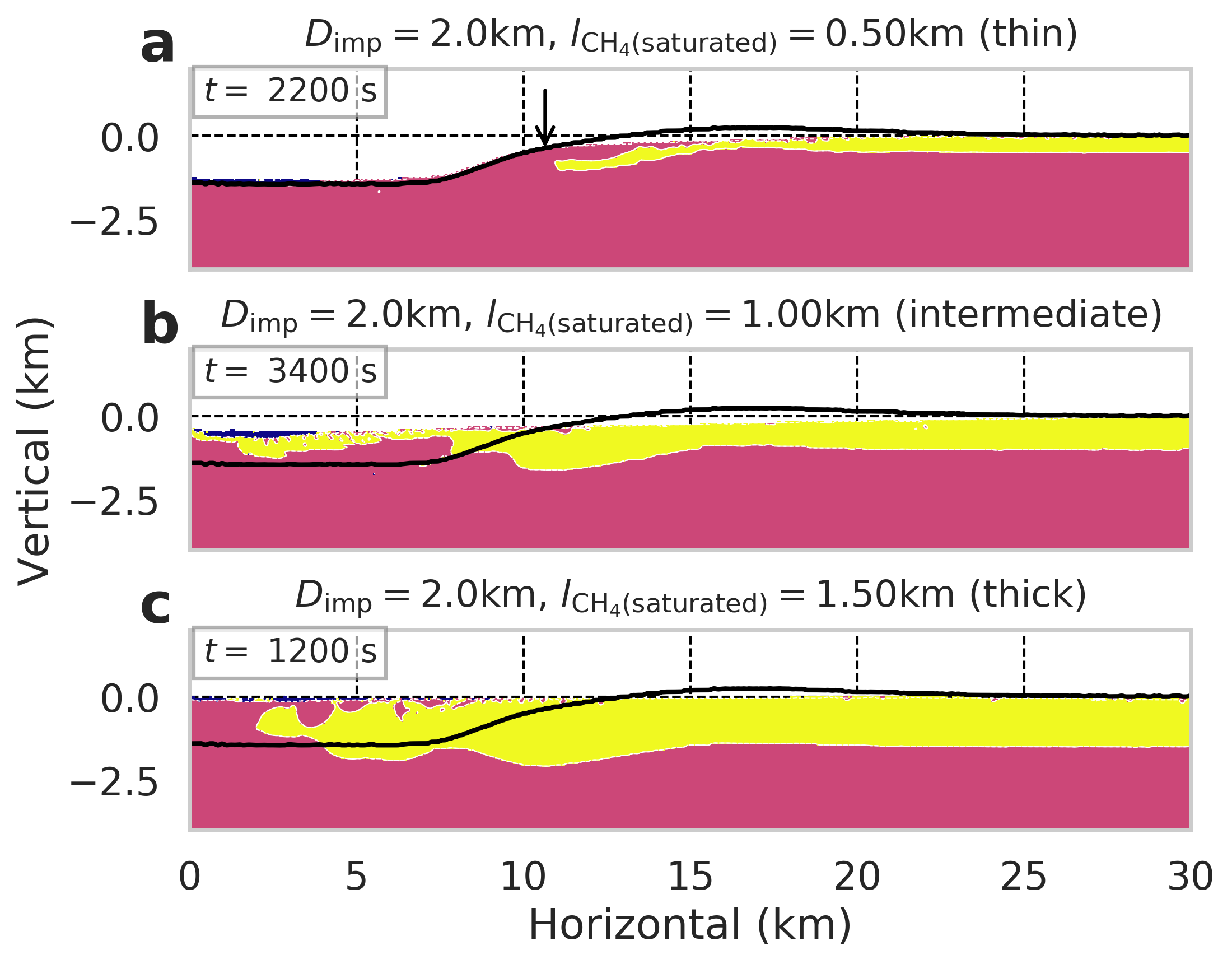}
\caption{
Same viewing as Fig. \ref{fig:dimp5km}, but for the case of $D_{\rm imp}$ = 2 km. The thickness of the methane-saturated layer is (a) 0.5 km, (b) 1 km, and (c) 1.5 km. 
\label{fig:dimp2km}}
\end{figure}

Next, we show the results for different impactor sizes with various methane-saturated layer thicknesses. 
In the same way of Figure \ref{fig:dimp5km}, we illustrate the results of $D_{\rm imp}$= 3 km in Figure \ref{fig:dimp3km} and $D_{\rm imp}$ = 2 km in Figure \ref{fig:dimp2km}. 
To examine these results, we categorize the methane-saturated layer thickness into three groups: thin, intermediate, and thick.

In the case of a “thin” methane-saturated layer ($\lesssim25$\% of the impactor diameter), the impact crater cavity is well preserved (Figs. \ref{fig:thin}, \ref{fig:dimp5km}a, \ref{fig:dimp3km}a, and \ref{fig:dimp2km}a). 
However, the methane-saturated layer suppresses the crater rims even when the cavity is minimally affected (see black lines which represent a methane-clathrate basement with no overlaying saturated layer). 
In the case of the crater with a methane-saturated layer, the surface profile beyond the crater wall is too subtle to define the highest point as we see in the case without it (see black lines in Figs. \ref{fig:dimp5km}a, \ref{fig:dimp3km}a, and \ref{fig:dimp2km}a).
Thus, we define the diameter of craters formed in a target with a methane-saturated layer based on the location of the edge of the wall (e.g., the slope becomes nearly flat as indicated by arrows in Figs.  \ref{fig:dimp5km}a, \ref{fig:dimp3km}a, and \ref{fig:dimp2km}a). 
The “obvious” depression becomes smaller than that of the non-saturated-layer case (see first paragraph in Section \ref{sec:disc}). 

As the thickness of the methane-saturated layer increases, the crater is filled in shortly after formation, and the crater becomes less recognizable. 
In the case that the methane-saturated layer is "intermediate" in thickness ($\sim$25--40\% of the impactor diameter), there is no crater rim or “obvious” depression, just a subtle depression (see 1.5 km thick for $D_{\rm imp}$ = 5 km (Fig. \ref{fig:dimp5km}b), 1.5 km for  $D_{\rm imp}$ = 3 km (Figs. \ref{fig:dimp3km}b), and 1 km thick for $D_{\rm imp}$ = 2 km (Fig. \ref{fig:dimp2km}b)). 
While it is difficult to define the radii of these craters, the resulting depressions appear wider than craters formed by comparable impacts into non-saturated icy targets (see black solid lines in Figs.  \ref{fig:dimp5km}b, \ref{fig:dimp3km}b, and \ref{fig:dimp2km}b). 
Note that the numerical resolution of 50 m or 100 m limits the confidence range in the horizontal direction up to 35 km or 70 km, respectively.  

If the methane-saturated layer is “thick” (thicker than $\sim$ 40--50\% of the impactor diameter), the resulting crater is completely filled by the methane-saturated ice. 
As we see in Figure \ref{fig:thick}, the methane-saturated layer fills in the depression, essentially erasing the impact crater. 
We find that the target will not preserve any crater in the case of a thick methane-saturated layer 
(i.e., 2 km for $D_{\rm imp}$ = 5 km (Fig. \ref{fig:dimp5km}c), 2 km for $D_{\rm imp}$ = 3 km (Fig. \ref{fig:dimp3km}c), and 1.5 km for $D_{\rm imp}$ = 2 km (Fig. \ref{fig:dimp2km}c)).

\section{Discussion} \label{sec:disc}
\begin{figure}
\plotone{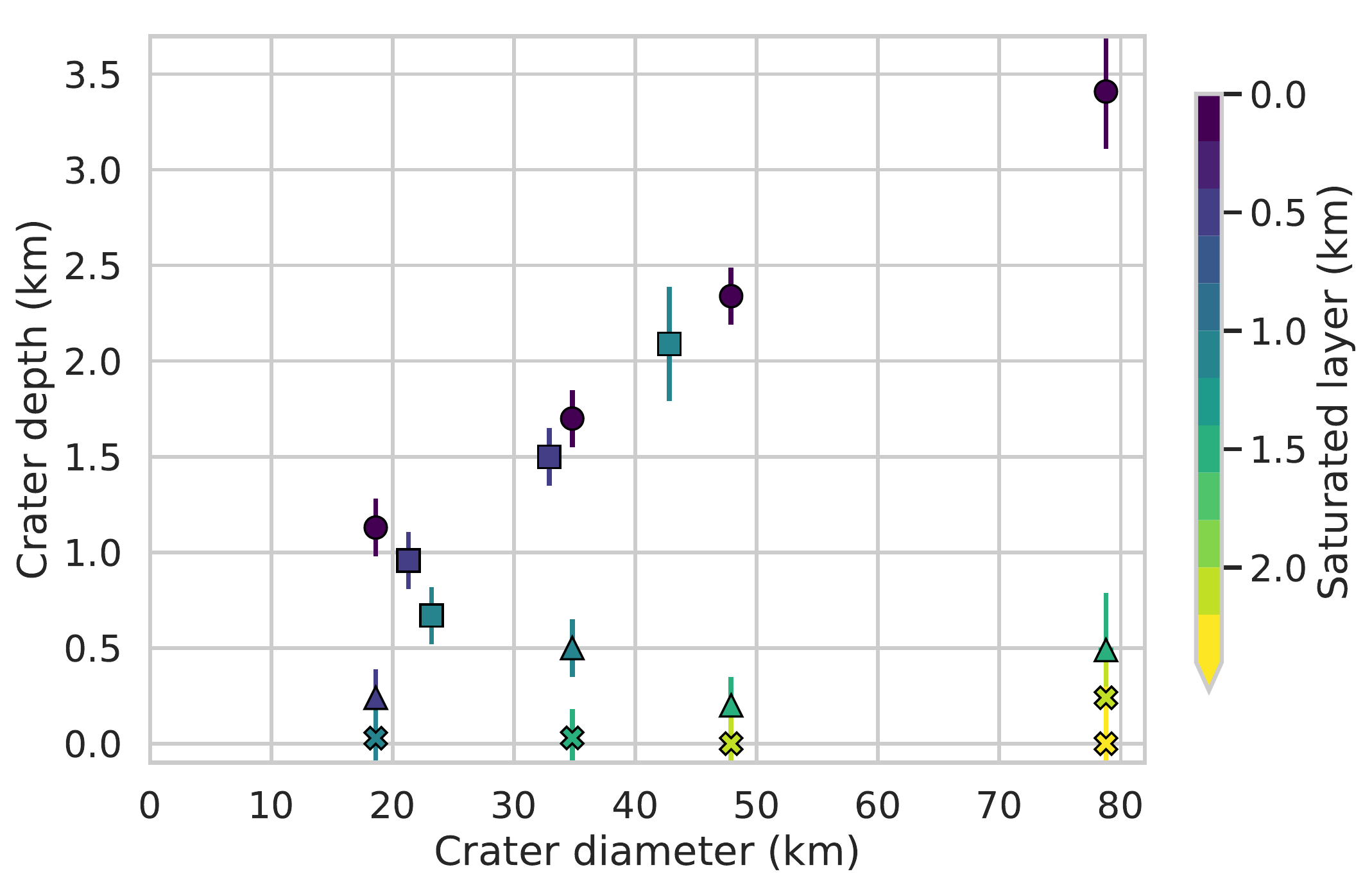}
\caption{
Crater depth as a function of crater diameter. 
Color indicates the thickness of the methane-saturated layer. 
Each symbol represents a different topographic expression of the crater; 
circle for an “obvious” cavity with rim, square for an “obvious” cavity without rim, triangle for depression, and cross for an flat surface (see also Table \ref{tab:crater}). 
The vertical error bar corresponds to the uncertainty of 6 cells. 
\label{fig:sum}}
\end{figure}

We first modeled impacts into methane clathrate, which is stronger than water ice. 
Our impact simulation results show that craters formed in methane-clathrate targets have a slightly smaller diameter and depth than those in water-ice targets.
This is because the stronger methane clathrate inhibits crater excavation but also limits crater collapse (see Fig. \ref{fig:d5_time} and supplemental movies \ref{sup:ch4} and \ref {sup:h2o}). 
We next simulate impacts into a weak methane-saturated layer over a strong methane-clathrate basement. 
Our results show that the morphology of these impact craters depends on the thickness of the methane-saturated layer.
Here, we examine the relationship between the diameter and depth of the resultant impact craters and the thickness of the methane-saturated layer. 
We summarize our results for the impact crater's diameter, depth, and morphology for a variety of different starting conditions in Table \ref{tab:crater}. 
Figure \ref{fig:sum} shows the diameter and the depth of the crater cavity, colored by the thickness of the methane-saturated layer. 
Even if a crater forms an obvious depression, the methane-saturated material at least partially fills it in, resulting in a shallower crater than a similar impact into a non-methane-saturated target. 
Since it is difficult to define a diameter without any obvious rim (i.e., the intermediate and thick methane-saturated layer cases), 
we plot them at the diameters of their respective non-saturated layer cases (triangles and cross symbols in Fig. \ref{fig:sum}). 

Thus, we find that a methane-saturated layer may help to explain the infrequency of craters in the lowlands of Titan, as suggested by \citet{Neish:2014}. 
Impacts into lowland regions with a thick methane-saturated layer (thickness $>$ 40--50 \% of the impactor diameter) will not result in any recognizable craters from an orbital platform. 
Impacts into thinner methane-saturated layers, as might be expected at intermediate elevations or as Titan's surface dried out (see square symbols in Fig. \ref{fig:sum}), would result in shallow, poorly preserved crater structures that might resemble the “possible” or “probable” craters observed on Titan. 
For smaller craters, this puts a rather strict limit on the maximum thickness of the methane-saturated layer -- for instance, an 18-km diameter crater can only be preserved if the methane-saturated layer is less than 0.5 km thick. 
Craters formed in regions with no (or at most a superficial) methane-saturated layer are the most reasonable candidates for the “certain” craters observed on Titan, as they will have obvious crater cavities and rims. 
As such, the variation in methane-saturated layer thickness may help to explain the various morphologies of craters observed on Titan. 
However, there is a significant difference in the crater depth between the thin and intermediate methane-saturated layer cases (see squares and triangles in Fig. \ref{fig:sum}). 
Craters go from several kilometers in depth to essentially flat, with few cases in between. 
Thus, the methane-saturated layer cannot easily account for the large number of craters on Titan observed with moderate depths \citep[0.1 to 0.8 km,][]{Hedgepeth:2020}. 
Some amount of erosion by aeolian infill and/or fluvial incision seems necessary to explain these observations.
We also recognize that there are some lowland regions with craters on Titan, notably in Xanadu and Senkyo \citep{Wood:2010,Buratti:2012}, and the largest crater on Titan, Menrva, is located in-between the lowland and highland regions \citep{Crosta:2021}. As such, there are some exceptional lowland regions that contains craters. These may have been formed in a “drier” part of Titan's history.

Most impact craters on Titan have been identified by their morphology from orbital observations, i.e., a bright circular rim seen in Cassini RADAR images \citep[e.g.,][]{Wood:2010,Hedgepeth:2020}. 
Traditionally, topography is the gold standard by which impact craters are identified on other worlds, but topographic data is limited on Titan. 
Our work, however, suggests that evidence of some of Titan's impacts may come from observed differences in surface composition. 
We found that impacts into intermediate and thick methane-saturated layers will result in significant differences in the composition of material exposed on the surface. Specifically, the methane-clathrate layer that was buried beneath the methane-saturated layer will rise to the surface (Figs. \ref{fig:dimp5km}b,c, \ref{fig:dimp3km}b,c, and \ref{fig:dimp2km}c).
We may be able to observe such compositional differences even in the absence of clear morphologic features such as a circular depression or rim, given an in situ asset like Dragonfly. 
The Dragonfly spacecraft will explore the Titan surface, especially around the Selk impact crater \citep{Barnes:2021,Lorenz:2021}.
If such compositional variations are detected, we can estimate the size of impactor that formed Selk and the methane-saturated layer thickness using our numerical results.

Finally, we note that there are some uncertainties in our approach that could affect our model results. The strength of the target, which is key to determining the crater morphology, is not well constrained.
We had to model the methane-saturated layer by adapting a lower frictional coefficient and limiting strength compared to water ice, as there is no experimental or observational data for methane-saturated sediments on Titan.
Changes in our assumptions regarding the strength of the methane-saturated layer could change the topographic expression, depth, and diameter of the modeled craters. 
For example, if the methane-saturated layer is stronger than what we used in our model, the thickness of the methane-saturated layer needed to erase the crater cavity increases.
However, it would not change the essential behaviour of the methane-saturated layer; the weak sediment layer will still fill in the crater.
Experimental data of \citet{Durham:2003} was used to develop the strength model of methane clathrate, which is stronger than water ice.
However, we use the equation of state (EOS) of water ice to represent the methane-clathrate basement and the methane-saturated layer.
Although the influence of the EOS on the topographic expression of crater is unknown, it likely affects the amount of melt produced during the impact and its freezing time after crater formation ceases. 
If the methane clathrate acts like porous water ice, it could form a larger melt pool than non-porous water ice \citep{Wunnemann:2008}.

We also wish to point out that although impacts into “thin” methane-saturated layers resulted in craters with an obvious depression, our model craters are still deeper than most observed Titan craters \citep{Hedgepeth:2020}. 
The reason for this is unclear, but post-impact exogenic processes could change the morphology and depth of the initial cavity.
Fluvial erosion and aeolian infill will both make the crater shallower. 
As one example, \cite{Neish:2016} estimate that a 1-km-deep, 40-km-diameter crater would lose its surface expression over $\sim 500$ Myr.
Fluvial erosion would also affect the crater rim height, possibly producing the lower rim heights observed on Titan compared to Ganymede \citep{Hedgepeth:2020}. 
Exogenic processing could more easily fill in the smaller depressions observed in the “intermediate” thick saturated layer, so the topographic expression of these craters will likely be erased more quickly. 
Alternatively, a higher temperature gradient in Titan’s interior could form a shallower crater, as was observed to be the case for Ganymede and Callisto \citep{Bjonnes:2018}, and reduce the crater rim height due to the overflow of warm ice \citep{Silber:2017}. 
In case of a higher temperature gradient than we consider in this work, an initial crater formed in methane-clathrate target would be shallower and more subtle. 
If the initial crater cavities have shallower depths, a thinner methane-saturated layer over the methane-clathrate basement would be sufficient to erase the crater topography.
Additional detailed analyses are needed to understand the role that temperature gradients play in the initial crater depth on Titan. 
This will influence how craters evolve over time and how easily a crater cavity is formed in the presence of a methane-saturated layer. 

\section{Conclusions} \label{sec:conc}
We performed simulations of impacts into a Titan crust that modeled as a methane-saturated layer over a methane-clathrate basement, which lies on top of a liquid water ocean. 
We first developed a strength model for methane clathrate, which is stronger than that of water ice \citep{Durham:2003}. 
Although we made some assumptions in our strength model of methane clathrate, craters formed in a methane-clathrate target appear slightly shallower and smaller in diameter than those formed in the pure water-ice target. 
Further experimental data of methane clathrate is essential to improve the strength model of methane clathrate; those data acquired under Titan conditions will be especially helpful for simulating Titan crater formation.
We next investigated the influence of a methane-saturated layer on top of the methane-clathrate basement. 
As the methane-saturated layer thickness increases, it has a larger influence on the final morphology of an impact crater; 
a thin layer of methane-saturated ice ($<$ 25\% of the impactor diameter) eliminates the crater rim but leaves the cavity relatively unchanged,
whereas a thick layer of methane-saturated ice ($>$ 40--50\% of the impactor diameter) completely erases the crater cavity. 
A thick methane-saturated layer supports the hypothesis for the dearth of lowland craters on Titan \citep{Neish:2014}, if liquids do indeed pool in these regions of low elevation. 
A lack of observed craters in regions of methane saturation on Titan will affect its inferred surface age, making the surface appear younger than it truly is.

\begin{acknowledgments}
We gratefully acknowledge the developers of iSALE-2D, including Gareth Collins, Kai W\"{u}nnemann, Dirk Elbeshausen, Tom Davison, Boris Ivanov, and Jay Melosh (\url{http://www.isale-code.de}).
We also thank Tom Davison, the developers of pySALEPlot tool, which helped us making some plots in this work.
This research was supported in part through computational resources provided by Information Technology at Purdue, West Lafayette, Indiana.
This work was supported by Cassini Data Analysis Program grant 80NSSC20K0382.
We thank the two anonymous reviewers for their helpful feedback.
\end{acknowledgments}


\bibliography{./titan_impacts.bbl}{}
\bibliographystyle{aasjournal}

\begin{deluxetable*}{lccc}
\tablecaption{iSALE input parameters \label{tab:input}}
\tablewidth{0pt}
\tablehead{
\colhead{Description} & \colhead{Methane clathrate} & \colhead{Water ice$^a$} & \colhead{Methane-saturated ice}}
\startdata
Equation of state & \multicolumn3c{Tillotson, H$_2$O}  \\
Thermal softening parameter & 0.8$^b$ & 1.2 & 1.2 \\
Cohesion, undamaged (MPa) & 10 & 10 & 10 \\
Cohesion, damaged (MPa) & 0.01 & 0.01 & 0.01 \\
Frictional coefficient, damaged & 2.0 & 2.0 & 1.0$^d$ \\
Frictional coefficient, damaged & 0.6 & 0.6 & 0.01$^d$ \\
Limiting strength, undamaged (MPa) & 2.2$^c$ & 0.11 & 0.04$^d$ \\
Limiting strength, damaged (MPa) & 2.2$^c$ & 0.11 & 0.04$^d$ \\
\enddata
\tablecomments{Parameters are the same as that of water ice$^a$ unless otherwise mentioned.}
\tablenotetext{a}{\cite{Bray:2014, Silber:2017}}
\tablenotetext{b}{see text}
\tablenotetext{c}{Representing 20 times stronger methane clathrate than water ice \citep{Durham:2003}}
\tablenotetext{d}{Representing a weak water-ice sediment \citep{Collins:2005}}
\end{deluxetable*}

\begin{deluxetable*}{cccclc}
\tablecaption{Crater diameter and depth \label{tab:crater}}
\tablewidth{0pt}
\tablehead{
\colhead{Impactor} & \colhead{Methane-saturated} & \colhead{Crater} & \colhead{Crater} 
& \colhead{Crater morphology} & \colhead{Symbols} \\
\colhead{diameter} & \colhead{layer thickness} & \colhead{diameter} & \colhead{depth} & 
\colhead{} & \colhead{ in Figure \ref{fig:sum}} 
}
\startdata
5 & 0 & 78.8 & 3.4 & “obvious” cavity with rim & $\bigcirc$  \\
5 & 1 & 42.8 & 2.0 & “obvious” cavity without rim  & $\square$  \\
5 & 1.5 & $>$78.8 & 0.49 & depression & $\triangle$  \\
5 & 2 & NA & NA & flat surface & $\times$  \\
3 & 0 & 47.9 & 2.3 & “obvious” cavity with rim  & $\bigcirc$  \\
3 & 0.5 & 32.9 & 1.5 & “obvious” cavity without rim  & $\square$  \\
3 & 1 & 23.2 & 0.67 & “obvious” cavity without rim  & $\square$  \\
3 & 1.5 & $>$47.9 & 0.49 & depression & $\triangle$  \\
3 & 2 & NA & NA & flat surface& $\times$  \\
2 & 0 & 34.8 & 1.7 & “obvious” cavity with rim  & $\bigcirc$  \\
2 & 0.5 & 21.3 & 0.96 & “obvious” cavity without rim  & $\square$  \\
2 & 1 & $>$21.3 & 0.5 & depression & $\triangle$  \\
2 & 2 & NA & NA & flat surface& $\times$  \\
1 & 0 & 18.6 & 1.1 & crater cavity with rim  & $\bigcirc$  \\
1 & 0.5 & $>$18.6 & 0.24 & depression & $\triangle$  \\
1 & 1 & NA & NA & flat surface & $\times$  \\
\enddata
\tablecomments{All units are in km.}
\end{deluxetable*}

\begin{figure}
\plotone{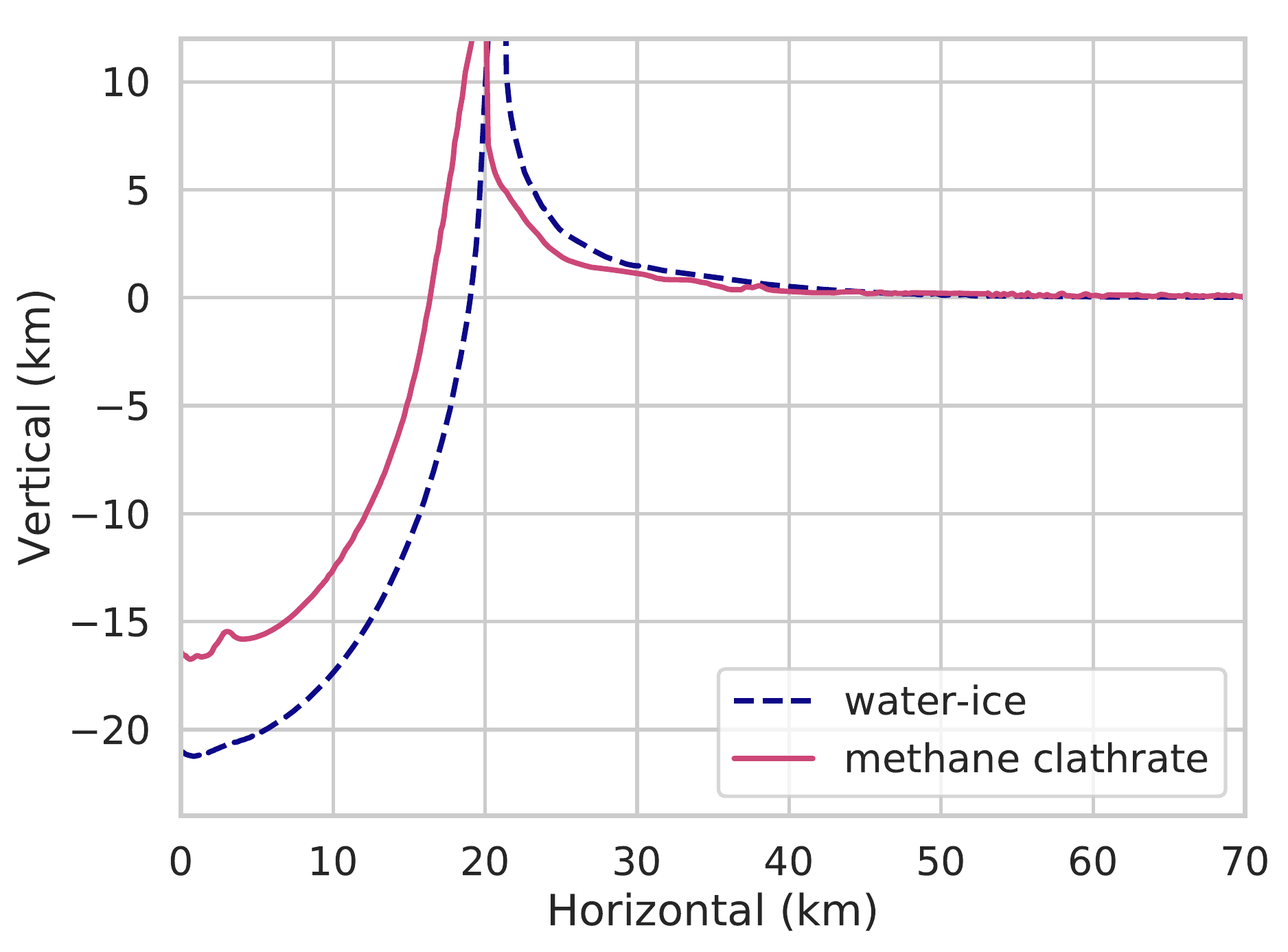}
\caption{
Surface profiles 60 s after the impact of a 5-km-diameter impactor into different targets. The blue dashed line depicts a water-ice basement, and the pink solid line depicts a methane-clathrate basement. 
\label{fig:d5_time}}
\end{figure}

\appendix
\section{Supplemental Movies} \label{sec:appendix}
\restartappendixnumbering

\begin{figure}
\begin{interactive}{animation}{ch4.mp4}
\plotone{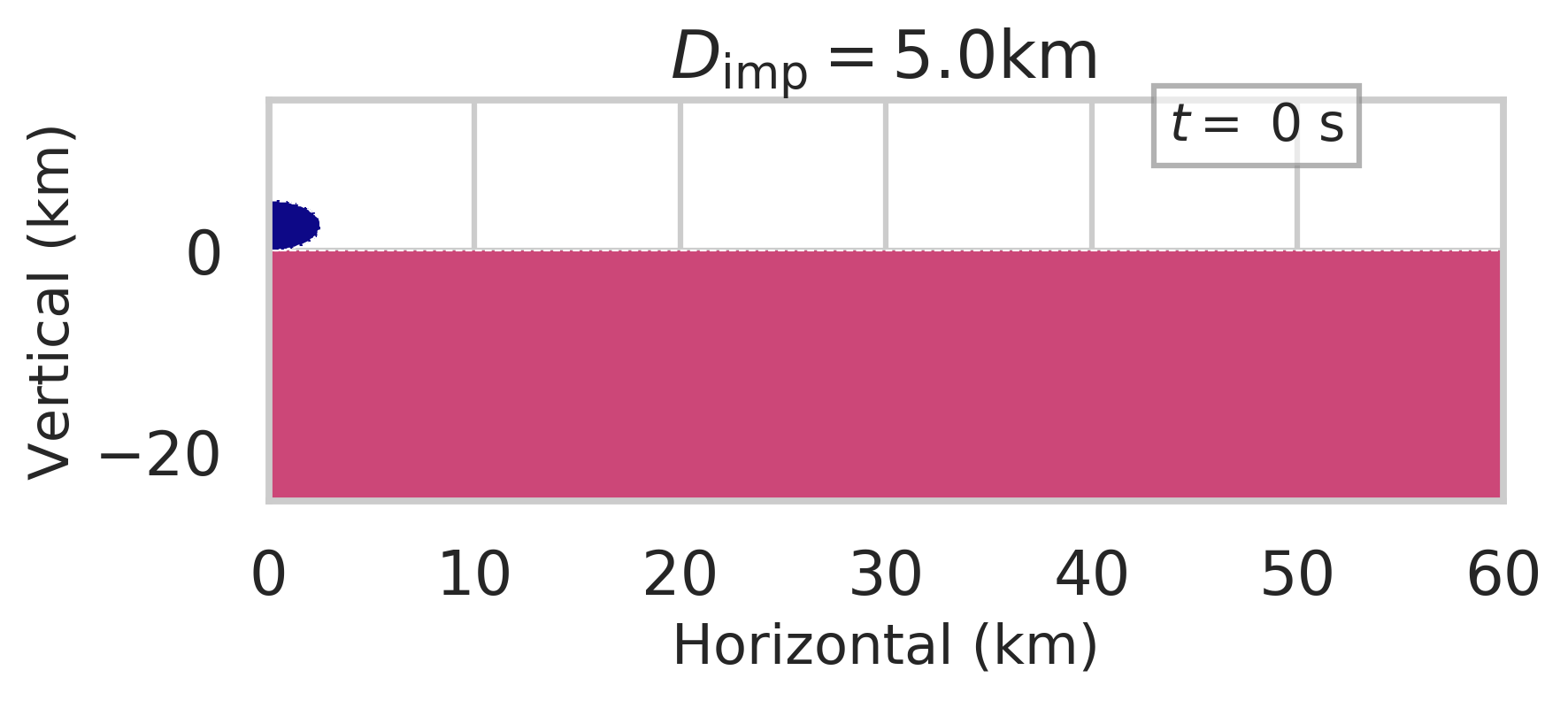}
\end{interactive}
\caption{The animation for the impact of a $D_{\rm imp}$ = 5 km impactor into a methane-clathrate basement. \label{sup:ch4}}
\end{figure}

\begin{figure}
\begin{interactive}{animation}{h2o.mp4}
\plotone{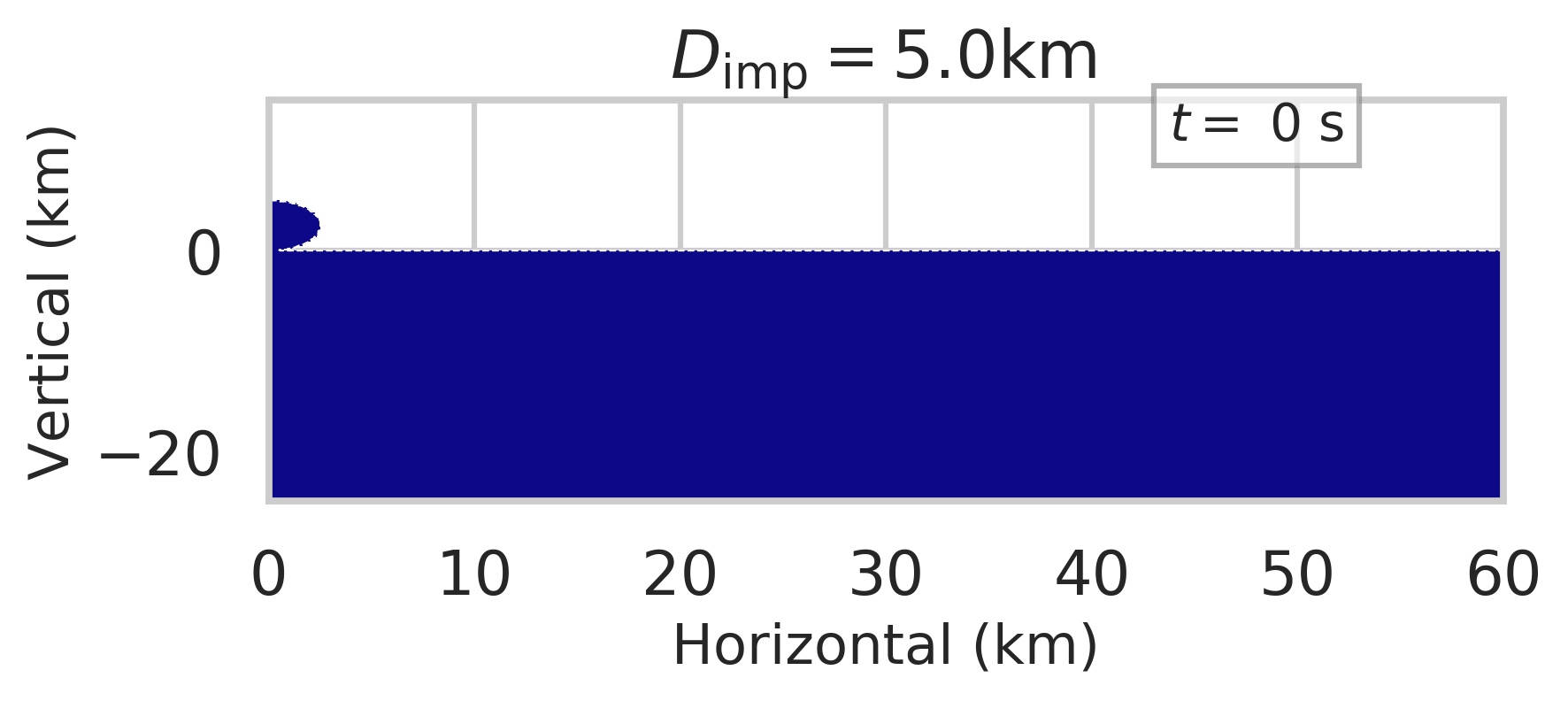}
\end{interactive}
\caption{Same viewing as Fig. \ref{sup:ch4}, but for a water-ice basement. \label{sup:h2o}}
\end{figure}

\begin{figure}
\begin{interactive}{animation}{ch4sthin.mp4}
\plotone{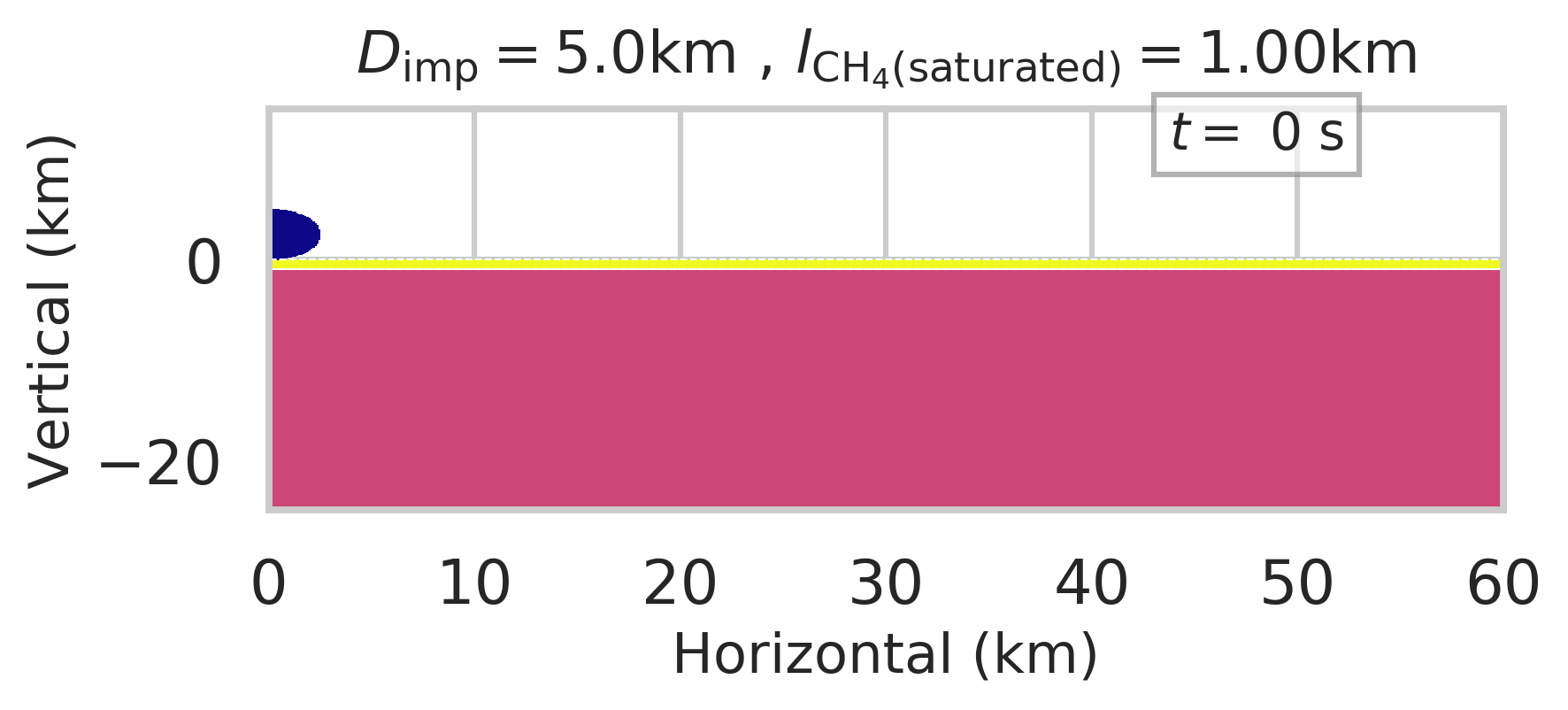}
\end{interactive}
\caption{Same viewing as Fig. \ref{sup:ch4}, but for a target composed of a 1 km methane-saturated layer over a methane-clathrate basement. \label{sup:ch4sthin}}
\end{figure}

\begin{figure}
\begin{interactive}{animation}{ch4smid.mp4}
\plotone{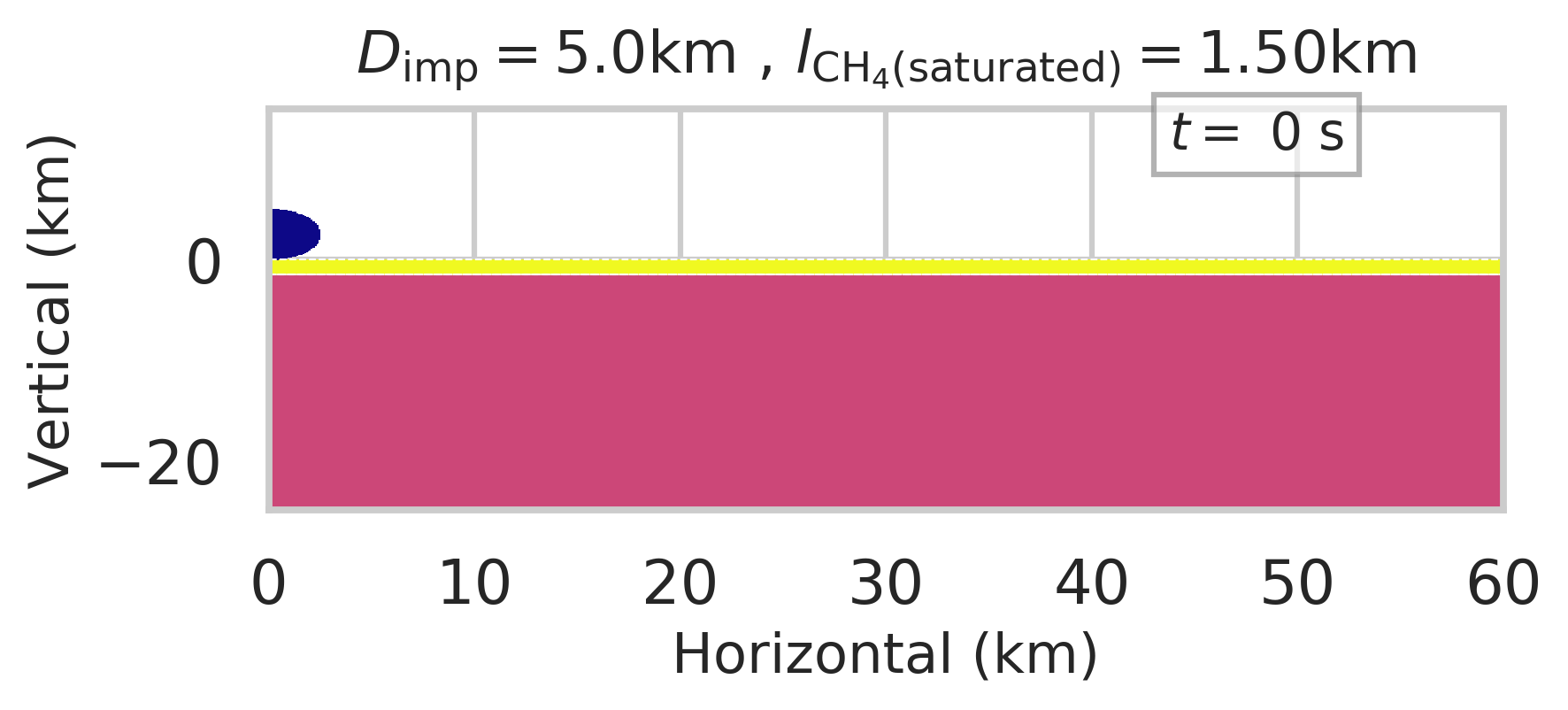}
\end{interactive}
\caption{Same viewing as Fig. \ref{sup:ch4}, but for a target composed of a 1.5 km methane-saturated layer over a methane-clathrate basement. \label{sup:ch4smid}}
\end{figure}

\begin{figure}
\begin{interactive}{animation}{ch4sthick00.mp4}
\plotone{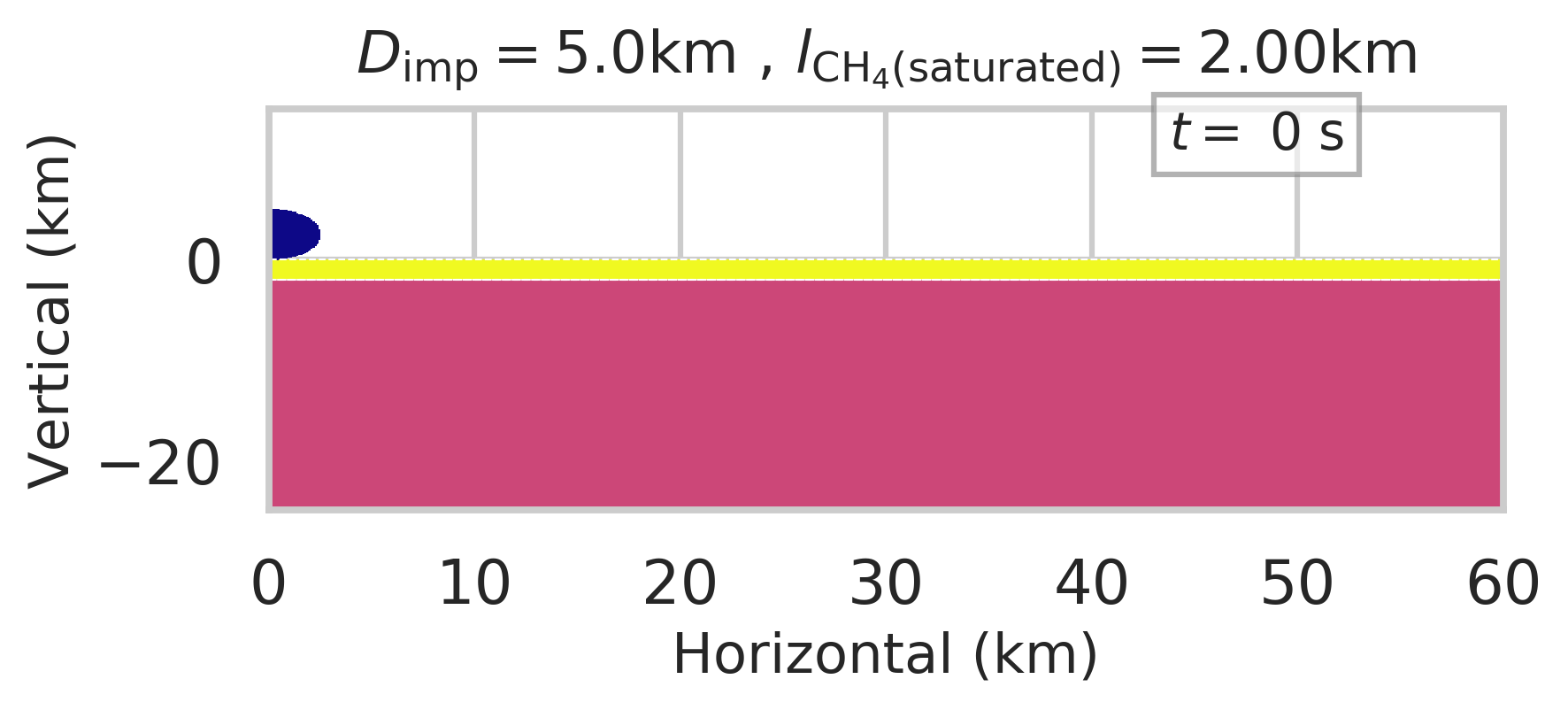}
\end{interactive}
\caption{Same viewing as Fig. \ref{sup:ch4}, but for a target composed of a 2 km methane-saturated layer over a methane-clathrate basement. \label{sup:ch4sthick}}
\end{figure}



\end{document}